\documentclass[aps,prb,twocolumn,superscriptaddress,longbibliography]{revtex4-2}
\usepackage{graphicx}
\usepackage{latexsym}
\usepackage{amssymb}
\usepackage{amsmath}
\usepackage{amsfonts}
\usepackage{upgreek}
\usepackage{float}
\usepackage{bm}
\usepackage{multirow}
\usepackage{color}
\usepackage[T1]{fontenc}
\usepackage{hyperref}
\usepackage{subfigure}
\usepackage{booktabs,tabularx}

\hypersetup{
colorlinks = true,
linkcolor = [rgb]{0.70,0.13,0.13},
citecolor = [rgb]{0.13,0.55,0.13},
urlcolor  = [rgb]{0.25, 0.41, 0.88}}
\newcommand{\ket}[1]{|#1\rangle}
\newcommand{\bra}[1]{\langle#1|}
\newcommand{\braket}[2]{\langle#1|      #2\rangle}

\begin{document}
\title{Dynamical signatures of the Yang-Lee edge singularity in non-Hermitian systems}
\author{Ming-Chu Lu}
\affiliation{College of Physics, Nanjing University of Aeronautics and Astronautics, Nanjing, 211106, China}

\author{Shun-Hui Shi}
\affiliation{College of Physics, Nanjing University of Aeronautics and Astronautics, Nanjing, 211106, China}

\author{Gaoyong Sun}
\thanks{Corresponding author: gysun@nuaa.edu.cn}
\affiliation{College of Physics, Nanjing University of Aeronautics and Astronautics, Nanjing, 211106, China}
\affiliation{Key Laboratory of Aerospace Information Materials and Physics (NUAA), MIIT, Nanjing 211106, China}

\begin{abstract}
The Yang-Lee edge singularity is an intriguing critical phenomenon characterized by nonunitary field theory.
However, its experimental realization for interacting many-body systems remains elusive.
We show that Yang-Lee edge singularities, regarded as many-body exceptional points, can be observed using both the self-normal and the associated-biorthogonal Loschmidt echoes, leveraging the advantages of nonunitary dynamics in non-Hermitian systems. The Loschmidt echoes are demonstrated to display unitary dynamics in the $\mathcal{PT}$-symmetric regime but exhibit nonunitary dynamics in the $\mathcal{PT}$ symmetry-broken regime, leading to a sharp change near an exceptional point. We hereby identify exceptional points in both the non-Hermitian transverse field Ising model and the Yang-Lee model, and determine the critical exponent that is consistent with nonunitary conformal field theory. This work provides a direct observation of Yang-Lee edge singularities in non-Hermitian many-body systems arising from nonunitary dynamics.

\end{abstract}

\maketitle

\section{Introduction}
Phase transitions and their universality classes are fundamental topics in physics \cite{Sachdev1999,vojta2003quantum}. 
The Ising model, a renowned benchmark example of phase transitions, plays a crucial role in understanding their nature \cite{Sachdev1999,vojta2003quantum}. 
To understand the nature of phase transitions and their mechanisms, Yang and Lee investigated the Ising model with an external magnetic field in the complex plane in two seminal papers \cite{yang1952statistical,lee1952statistical}.
In Lee-Yang theory, phase transitions are interpreted through the zeros of the partition function, known as Lee-Yang zeros \cite{yang1952statistical,lee1952statistical}.
For the Ising model, the Lee-Yang zeros are purely imaginary, according to the Lee-Yang theorem \cite{yang1952statistical,lee1952statistical}.
Above the critical temperature, the Lee-Yang zeros concentrate along a line in the thermodynamic limit, with a gap in their distribution \cite{fisher1978yang}. 
The edges of this gap are referred to as Yang-Lee edge singularities \cite{fisher1978yang}.

Yang-Lee edge singularities are fascinating nonunitary critical points with simple unconventional universality classes that depend solely on the dimensionality and can be described by a nonunitary CFT \cite{fisher1978yang,kortman1971density,cardy1985conformal}.
However, the experimental realization of Yang-Lee edge singularities is particularly challenging as they appear in complex physical parameters.
One proposal to realize the effect of complex physical parameters is to use a probe spin coupled to a system, where the Lee-Yang zeros correspond to the zeros of spin coherences \cite{wei2012lee}. 
Recently, the Lee-Yang zeros have been observed by directly measuring the coherence of a probe spin \cite{peng2015experimental}, in contrast to previous indirect detection experiment \cite{binek1998density,binek2001yang}.
Yang-Lee edge singularities, regarded as exceptional points \cite{uzelac1979yang,von1991critical,bianchini2014entanglement,sanno2022engineering,yamada2022matrix,wei2023tensor}, represent nonunitary criticality in non-Hermitian systems.
Thanks to the development of non-Hermitian physics \cite{bergholtz2021exceptional,ashida2020non,lee2016anomalous,yao2018edge,kunst2018biorthogonal,gong2018topological,yokomizo2019non,yang2020non,okuma2020topological,zhang2020correspondence,borgnia2020non}, 
Yang-Lee edge singularities have recently been observed in non-Hermitian quantum systems \cite{gao2024experimental,lan2024experimental}. 
This observation, in accordance with quantum-classical correspondence \cite{matsumoto2022embedding}, has led to the derivation of critical exponents for the one-dimensional classical Ising model.
Despite proposals for realizing Yang-Lee edge singularities in one-dimensional quantum many-body systems, such as utilizing Kibble-Zurek scaling \cite{yin2017kibble}, exploiting the quantum coherence of a probe spin \cite{wei2017probing}, and investigating dissipative Rydberg atomic arrays \cite{shen2023proposal}, achieving the realization of the critical points and critical exponents of Yang-Lee edge singularities remains elusive.

The Loschmidt echo (LE), a key concept in quantum information theory, quantifies information loss during the time evolution of a quantum system, serving as a probe to understand its dynamical behavior.
Specifically, the LE can exhibit an enhanced decay behavior \cite{quan2006decay} near the critical point, which enables the application of finite-size scaling theory \cite{fisher1972scaling,fisher1974renormalization}.
Motivated by this, a dynamical scaling theory for LE has been established \cite{hwang2019universality}, providing a powerful tool for characterizing equilibrium phase transitions \cite{hwang2019universality,tang2023dynamical}.
Recently, the scaling theory of LE was extended to non-Hermitian systems \cite{tang2022dynamical} by introducing a biorthogonal-unitary time evolution, establishing it as a valid observable for detecting phase transitions in the real energy regime. 
We must note that the successful application of dynamical scaling of LE in non-Hermitian systems with real energies is a fortunate exception, considering that time evolution is typically nonunitary in non-Hermitian systems \cite{orus2008infinite}. 
This nonunitary time evolution can lead to distinct dynamics across different quantum phases \cite{shen2023proposal}. 
Furthermore, it enables the detection of phase transitions in Hermitian systems even in the presence of non-Hermitian perturbations \cite{zhang2021quantum,zhang2021probing,liang2023probing}.
Consequently, it is not surprising that the biorthogonal-unitary time evolution developed in Ref.\cite{tang2022dynamical} cannot describe the phases in the complex energy regime. Recently, a biorthogonal time evolution approach in the associated bases \cite{brody2013biorthogonal} has been introduced to handle nonunitary dynamics in the complex energy regime \cite{jing2024biorthogonal}, revealing interesting biorthogonal dynamical quantum phase transitions.

In this paper, we will use the LEs, which are readily achievable in experiments \cite{jurcevic2017direct,wang2019simulating,tonielli2020ramsey}, to detect the critical phenomena \cite{xiao2019observation} and critical exponents of Yang-Lee edge singularities \cite{gao2024experimental} understood as exceptional points.
We implement biorthogonal LEs in the associated bases and study their responses in both the non-Hermitian transverse field Ising model and the Yang-Lee model. 
Our analysis reveals that the Yang-Lee edge singularity can be characterized by the dynamical signatures of different phases, offering a direct observation of Yang-Lee edge singularities in non-Hermitian many-body systems.

This paper is organized as follows. 
In Sec.\ref{sec:Loschmidtecho}, we introduce the concepts of the self-normal Loschmidt echo and the associated-biorthogonal Loschmidt echo.
In Sec.\ref{sec:IsingModel}, we present the results for the non-Hermitian transversed field Ising model.
In Sec.\ref{sec:YangLeeModel}, we introduce and analyze phase transitions in the Yang-Lee model.
In Sec.\ref{sec:conclusion}, we summarize our results.

\begin{figure}[t]
	\centering
	\includegraphics[width=8.0cm]{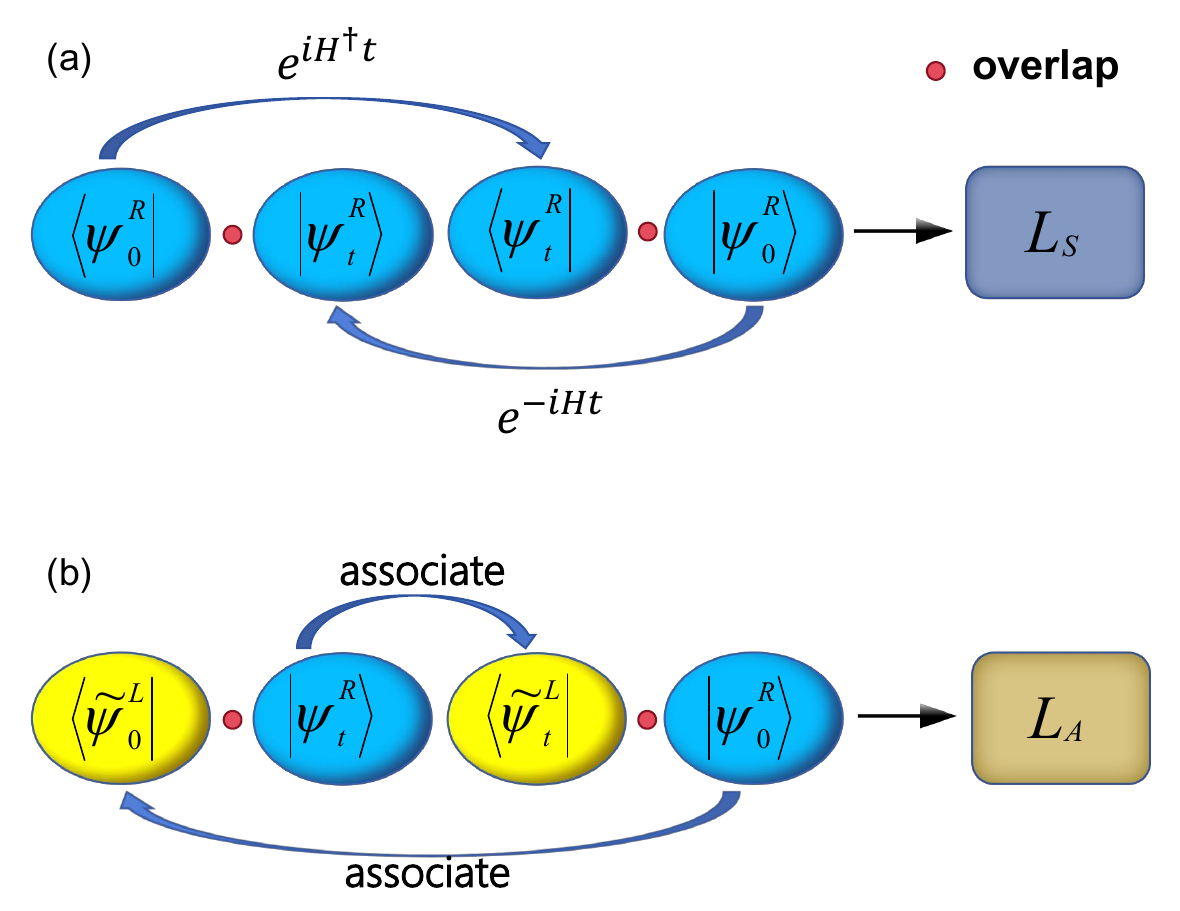}
	\caption{Schematic diagram of Loschmidt echoes. (a) The self-normal Loschmidt echo $L_{S}$, (b) The associated biorthogonal Loschmidt echo $L_{A}$. 
	Blue symbols represents the right eigenstates, and yellow symbols denotes their associated states.
	}
	\label{fig:LE}
\end{figure}

\section{Loschmidt echoes}
\label{sec:Loschmidtecho}
In a Hermitian system, the LE measures the probability that the time-evolved state returns to the initial state, which is defined as \cite{quan2006decay},
\begin{equation}
	L(\lambda_f,\lambda_i,t)=|\braket{\psi_{n}(\lambda_i)}{\psi_{n}(\lambda_f,\lambda_i,t)}|^2,
	\label{eq:LE}
\end{equation}
where $\ket{\psi_{n}(\lambda_i)}$ denotes the initial state, which is the $n$th eigenstate of the Hamiltonian $H(\lambda)$ at parameter $\lambda_{i}$,
and $\ket{\psi_{n}(\lambda_f,\lambda_i,t)}$ represents the time-evolved state of $\ket{\psi_{n}(\lambda_i)}$ following a quench from $H(\lambda_i)$ to $H(\lambda_f)$.

For non-Hermitian systems, where the Hamiltonian is non-Hermitian, $H(\lambda) \neq H^\dagger(\lambda)$, we obtain two sets of eigenstates, namely the right eigenstates $\ket{\psi^{R}_{n}(\lambda)}$ and the left eigenstates $\ket{\psi^{L}_{n}(\lambda)}$, dubbed as biorthogonal eigenvectors, from the eigenvalue equations \cite{brody2013biorthogonal},
\begin{align}
	H(\lambda)\ket {\psi^{R}_{n}(\lambda)}=\epsilon _{n}(\lambda)\ket{\psi^{R}_{n}(\lambda)}, \label{eq:eigenequationR} \\ 
	H^\dagger(\lambda)\ket {\psi^{L}_{n}(\lambda)}=\epsilon^\ast  _{n}(\lambda)\ket{\psi^{L}_{n}(\lambda)}, 
	\label{eq:eigenequationL}
\end{align}
where $\epsilon_{n}(\lambda)$ and $\epsilon^\ast_{n}(\lambda)$ represent the $n$th right and left eigenvalues.
It is known that biorthogonal eigenvectors satisfy both the completeness relation \cite{brody2013biorthogonal},
\begin{align}
\sum_{n}\ket{\psi_{n}^{R}(\lambda)}\bra{\psi_{n}^{L}(\lambda)}=1,
\end{align}
and the biorthogonal relation \cite{brody2013biorthogonal},
\begin{align}
\braket{\psi^{L}_{m}(\lambda)}{\psi^{R}_{n}(\lambda)}=\delta _{m,n}. 
\end{align}

The time-evolved states of $\ket{\psi^{R}_{0}(\lambda_i)}$ and $\ket{\psi^{L}_{0}(\lambda_i)}$ in biorthogonal bases under a quench from $\lambda_i$ to $\lambda_f$ are given by \cite{tang2022dynamical},
\begin{align}
       \ket{\psi^{R}_{0}(\lambda_f,\lambda_i,t)}=e^{-iH_ft}\ket{\psi^{R}_{0}(\lambda_i)}, \label{eq:timeEvolveR}  \\
       \ket{\psi^{L}_{0}(\lambda_f,\lambda_i,t)}=e^{-iH_f^{\dagger}t}\ket{\psi^{L}_{0}(\lambda_i)}. 
       \label{eq:timeEvolveL}
\end{align}
We note that in our study, we focus on the perturbation of the system (a small quench $\delta \lambda=\lambda_f - \lambda_i$). 
The Hamiltonian used for the time evolution is given by,
\begin{equation}
	H(\lambda_f)=H(\lambda_i)+\delta H(\lambda).
	\label{eq:Hamiltonian}
\end{equation}
Here, $\delta H(\lambda)$ depends on changes of the parameter $\delta \lambda$.

Naturally, one may define two types of LEs: the self-normal LE [c.f. Figs.\ref{fig:LE}(a)],
\begin{align}
L_S(\lambda_f,\lambda_i,t)=\braket{\psi^{R}_{0}(\lambda_i)}{\psi^{R}_{0}(\lambda_f,\lambda_i,t)}\braket{\psi^{R}_{0}(\lambda_f,\lambda_i,t)}{\psi^{R}_{0}(\lambda_i)},
\label{eq:LES}
\end{align}
using only right eigenstates $\ket{\psi^{R}_{0}(\lambda)}$, and the biorthogonal LE,
\begin{align}
L_B(\lambda_f,\lambda_i,t)=\braket{\psi^{L}_{0}(\lambda_i)}{\psi^{R}_{0}(\lambda_f,\lambda_i,t)}\braket{\psi^L_{0}(\lambda_f,\lambda_i,t)}{\psi^{R}_{0}(\lambda_i)}.
\label{eq:LEB}
\end{align}
using both right eigenstates $\ket{\psi^{R}_{0}(\lambda)}$ and left eigenstates $\ket{\psi^{L}_{0}(\lambda)}$.

It is expected that the biorthogonal LE should describe the phase transition \cite{tang2022dynamical}, as a non-Hermitian Hamiltonian is diagonalized in biorthogonal bases. 
In fact, it does describe phase transitions in the real-energy regime \cite{tang2022dynamical}. 
However, in the complex-energy regime, the biorthogonal LE, $L_B(\lambda_f,\lambda_i,t)$, becomes meaningless. 
This is because the biorthogonal LE, as defined in Eq.(\ref{eq:LEB}), is unitary, whereas a non-Hermitian system evolves non-unitarily in the complex-energy regime.

In order to describe time evolution of the LE in the complex-energy regime within the biorthogonal bases, we introduce associated states as outlined in Ref.\cite{brody2013biorthogonal,jing2024biorthogonal}. 
For a right eigenstate $\ket{\psi^{R}_{0}}$, we consider it as a superposition of a series of eigenstates and define the associated state 
$\ket{\tilde{\psi}^{L}_{0}}$ using the operation,
\begin{align}
	\ket{\psi^{R}_{0}}=\sum_{n}c_n \ket{\phi_{n}^{R}}\rightarrow \ket{\tilde{\psi}^{L}_{0}}=\sum_{n}c_n\ket{\phi_{n}^{L}},
	\label{eq:associatedstate}
\end{align}
where the coefficient $c_n=\braket{\phi^{L}_{n}}{\psi^{R}_{0}}$ is the same as in $\bra{\tilde{\psi}^L_{0}}=\sum_{n} c_n^\ast \bra{\phi^L_{n}}$. 
Here, $\ket{\phi_{n}^{R}}$ and $\ket{\phi_{n}^{L}}$ forms a biorthogonal basis and $\ket{\tilde{\psi}^{L}}$ denotes the associated state.

The biorthogonal LE in the associated state bases, dubbed as the associated-biorthogonal LE,  is defined as [c.f. Figs.\ref{fig:LE}(b)]:
\begin{align}
L_A(\lambda_f,\lambda_i,t)=\braket{\tilde{\psi}^L_{0}(\lambda_i)}{\psi^R_{0}(\lambda_f,\lambda_i,t)}\braket{\tilde{\psi}^L_{0}(\lambda_f,\lambda_i,t)}{\psi^R_{0}(\lambda_i)}.
\label{eq:LEA}
\end{align}
It is important to note that the associated state $\ket{\tilde{\psi}^L(\lambda_f,\lambda_i,t)}$ is determined by equation (\ref{eq:associatedstate}), 
rather than equation (\ref{eq:timeEvolveL}).

\begin{figure}[t]
	\centering
	\includegraphics[width=9.1cm]{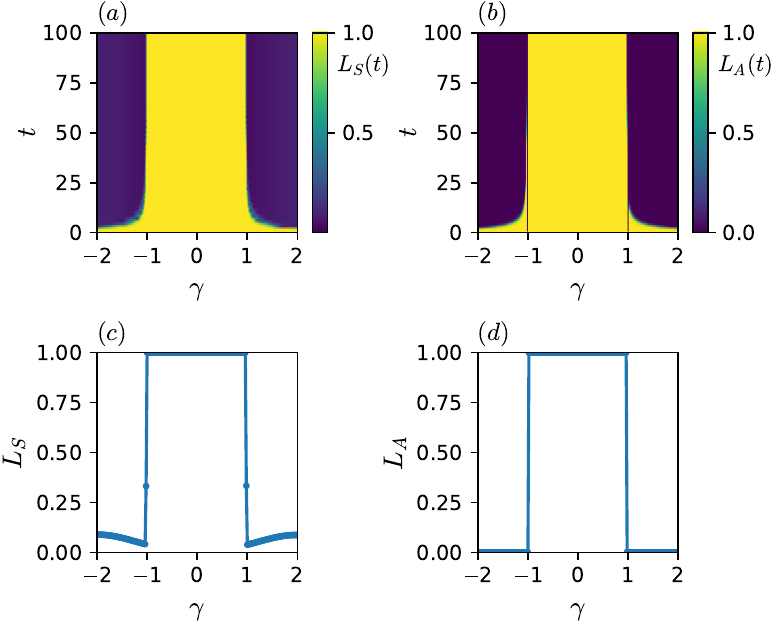}
	\caption{Time evolutions of LEs of the NHTI model. (a)(b) Self-normal and associated-biorthogonal LEs with respect to $\gamma$ and $t$. (c)(d) Self-normal and associated-biorthogonal LEs at $t=100$, where the LEs exhibit a sudden change at $\gamma=\pm1$, signaling a phase transition.
	}
	\label{fig:NHTI}
\end{figure}

\section{Non-Hermitian transversed field Ising model}
\label{sec:IsingModel}
In the realm of Hermitian systems, the transverse field Ising model serves as a standard benchmark for the study of phase transitions.
We now explore a one-dimensional non-Hermitian extension of this model, described by the Hamiltonian \cite{tang2022dynamical,zhang2020ising,sun2022biorthogonal,yang2022hidden,lu2024many}:
\begin{equation}
	H=-\sum_{j = 1}^{N}  J\sigma _{j}^{x}\sigma_{j+1}^{x}+\sum_{j = 1}^{N}\lambda(\sigma^{z}_{j}+i\gamma\sigma^{y}_{j}),
\label{eq:NHTI}
\end{equation}
where a complex field is introduced along the $y$-direction.
In this non-Hermitian transversed field Ising (NHTI) model, $J \geq 0$ denotes the strength of interaction between adjacent spins in the $x$-direction, defining the ferromagnetic coupling. 
$\lambda$ denotes the strength of the transverse field, whereas $\gamma$ introduces an additional field that quantifies the non-Hermiticity in the system.
Here, $\sigma_j^x, \sigma_j^y, \sigma_j^z$ denote the Pauli matrices along $x,y,z$-directions at lattice site $j$. 
Throughout the paper, we impose periodic boundary conditions such that $\sigma _{L+1}^{x,y,z}=\sigma _{1}^{x,y,z}$, where $N$ represents the length of the chain.

\begin{figure}[t]
	\centering
	\includegraphics[width=8.6cm]{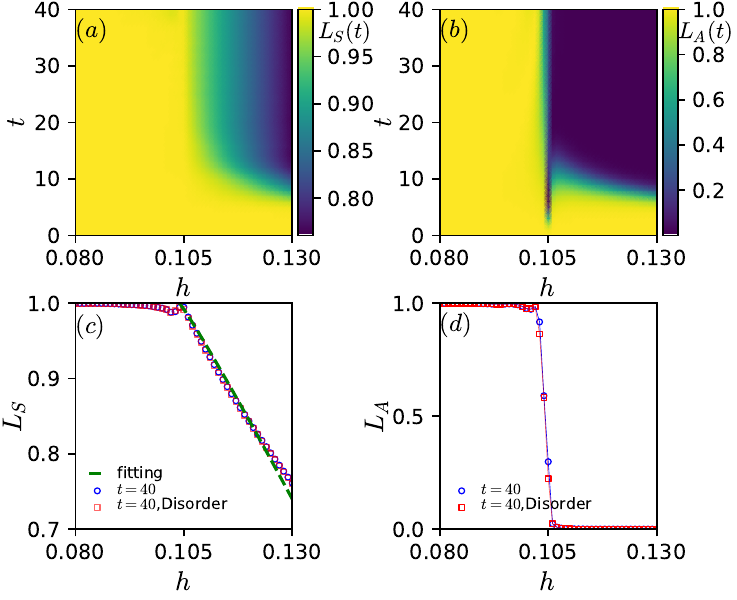}
	\caption{Time evolutions of LEs of the Yang-Lee model. (a)(b) Self-normal and associated-biorthogonal LEs for $N=12$, $\lambda=0.6$ with respect to $h$ and $t$. 
	Here, $\delta t=1$ and $\delta h=0.001$ are used. (c)(d) Self-normal and associated-biorthogonal LEs at $t=40$, where the LEs exhibit a change around $h_{N}= 0.104$, signaling a phase transition.
	Here, the blue circle and red square symbols represent the numerical data without and with disorders (by averaging over $10$ disorder pertubations), respectively. 
	The green dashed line in (c) denotes the linear fit, serving as a guide for the eye. 
	Through the critical point, $h_{N}= 0.104$, the system undergoes a transition from the $\mathcal{PT}$ symmetric phase to the $\mathcal{PT}$-symmetry broken phase. } 
	\label{fig:YLDynamics}
\end{figure}

The NHTI model can be solved exactly using a similarity transformation \cite{tang2022dynamical,zhang2020ising,sun2022biorthogonal,yang2022hidden,lu2024many}. It is known that this model exhibits an exceptional point at $\gamma = \pm 1$ \cite{tang2022dynamical,zhang2020ising,sun2022biorthogonal,yang2022hidden,lu2024many}. 
To determine if the LEs can capture this phase transition, we compute both the self-normal LE and the associated-biorthogonal LE in the NHTI model, 
where $\delta \lambda \equiv \delta \gamma$.
We use equation (\ref{eq:eigenequationR}) to obtain right eigenstates varying $\gamma$ from $-2$ to $2$ at $N=8$, $J=1$ and $\lambda=0.5$
and identify the right eigenstate with the lowest real part of the energy as the ground state, $\ket{\psi^R_{0}}$ during the simulations. The time-evolved state, $\ket{\psi^R_{0}(t)}$ is then obtained using Equation (\ref{eq:timeEvolveR}).
The associated states $\ket{\tilde{\psi}^L_{0}} , \ket{\tilde{\psi}^L_{0}(t)}$, used for calculating the associated-biorthogonal LE, are derived from Equation (\ref{eq:associatedstate}).

In the $\mathcal{PT}$ symmetric regime ($\left\lvert \gamma\right\rvert<1$), where the eigenvalues are real, we observe that both the self-normal LE and the associated-biorthogonal LE remain consistently close to one throughout the time evolution [c.f. Figs.\ref{fig:NHTI}]. 
In the $\mathcal{PT}$-symmetry broken regime, where the eigenvalues become complex and appear in conjugate pairs, we observe that both the self-normal LE and the associated-biorthogonal LE decay rapidly to zero during time evolution [c.f. Figs.\ref{fig:NHTI}]. A notable change in behavior is observed at the exceptional point, clearly indicating a phase transition at $\gamma = \pm 1$.

\begin{table}[t]
    \centering
    \renewcommand{\arraystretch}{1.0}
    \setlength{\tabcolsep}{7pt}
    \caption{Critical points of the Yang-Lee model for $\lambda$ ranging from $0.1$ to $0.9$ in the thermodynamic limit. Here $h_{\text{LE}}$, $h_{\text{MPRG}}$, $h_{\text{ED}}$ and $h_{\text{TNR}}$ represent the critical points $h_c$ obtained from the LEs, the matrix product renormalization group (MPRG) \cite{yamada2022matrix}, the exact diagonalization (ED) \cite{von1991critical} and the Loop tensor network renormalization (TNR) \cite{wei2023tensor} methods, respectively. }
	\label{tab:infiniteLE}
    \begin{tabular}{|c|c|c|c|c|c|}
    \hline
        $\lambda$ & $h_{\text{LE}} \text{(our work)}$ & $h_{\text{MPRG}}$ & $h_{\text{ED}}$ & $h_{\text{TNR}}$ \\ \hline
        0.1 & 0.6360 & 0.636 & 0.6366 & 0.6366\\ \hline
        0.2 & 0.4579 & 0.457 & 0.4585 & 0.4585\\ \hline
        0.3 & 0.3300 & 0.328 & 0.3300 & 0.3300\\ \hline
        0.4 & 0.2317 & 0.230 & 0.2320 & 0.2320\\ \hline
        0.5 & 0.1561 & 0.154 & 0.1562 & 0.1562\\ \hline
        0.6 & 0.0985 & 0.095 & 0.0981 & 0.0981\\ \hline
        0.7 & 0.0579 & 0.052 & 0.0548 & 0.0548\\ \hline
        0.8 & 0.0292 & 0.021 & 0.0247 & 0.0247\\ \hline
        0.9 & 0.0119 & -     & 0.0065 & 0.0065\\ \hline
    \end{tabular}
\end{table}

\section{Yang-Lee Model}
\label{sec:YangLeeModel}
The Yang-Lee model \cite{yang1952statistical,lee1952statistical}, proposed by Yang and Lee, extends the Ising model to explore the nature of phase transitions. It specifically investigates critical behaviors and phase transitions within the complex parameter plane. The Yang-Lee model is a non-Hermitian Ising model whose Hamiltonian is given by \cite{uzelac1979yang,von1991critical,bianchini2014entanglement,sanno2022engineering,yamada2022matrix,wei2023tensor}:
\begin{equation}
	H=-\sum_{j = 1}^{N}  (\lambda\sigma^{x}_{j}\sigma^{x}_{j+1}+ih\sigma^{x}_{j}+\sigma^{z}_{j}),
	\label{eq:YL}
\end{equation}
where $\lambda$ denotes the strength of the nearest-neighbor interaction along $x$-direction, and $h$ represents an external complex magnetic field, which can be interpreted as the contribution from dissipation acting on the system. 
This model undergoes a phase transition \cite{uzelac1979yang,von1991critical,bianchini2014entanglement,sanno2022engineering,yamada2022matrix,wei2023tensor}, with the critical point referred to as the Yang-Lee edge singularity.
In contrast to the NHTI model, the Yang-Lee model is a non-integrable model that cannot be solved exactly. Instead, it is analyzed using the framework of non-unitary conformal field theory (CFT) \cite{fisher1978yang,kortman1971density,cardy1985conformal}.
Additionally, numerical simulations \cite{uzelac1979yang,von1991critical,bianchini2014entanglement,sanno2022engineering,yamada2022matrix,wei2023tensor} have confirmed the results predicted by non-unitary CFT.
In the following, we will focus on the dynamical description of the Yang-Lee edge singularity through LEs, which are more easily realizable in experiments \cite{jurcevic2017direct,wang2019simulating,tonielli2020ramsey}.

In the Yang-Lee model, the Yang-Lee edge singularity is an exceptional point \cite{uzelac1979yang,von1991critical,bianchini2014entanglement,sanno2022engineering,yamada2022matrix,wei2023tensor}. 
Specifically, in the $\mathcal{PT}$ symmetric regime, the eigenvalues are real.
Conversely, in the $\mathcal{PT}$-symmetry broken regime, the eigenvalues become complex and appear in conjugate pairs. 
This should lead to distinct dynamical behaviors, as demonstrated in the NHTI model.
To evaluate the validity of this argument, we compute both the self-normal LE and the associated-biorthogonal LE with $h$ ranging from $0.08$ to $0.13$, using $\delta \lambda \equiv \delta h$ for $N=12$ at $\lambda=0.6$.  
The results of the LEs are presented in Figs.\ref{fig:YLDynamics}. We observe that, similar to the NHTI model, the associated-biorthogonal LE decays rapidly from an initial value of one to zero, while the self-normal LE exhibits an approximately linear decay. 
Moreover, the behavior of the LEs is examined for robustness against random perturbations of the form $i \epsilon \sigma^{x}_{j}$, where $\epsilon$ is uniformly distributed within the interval $[-\delta h, \delta h]$ [c.f.  Figs.\ref{fig:YLDynamics}(c) and (d)].
Critical points are determined by the intersection of the linear decay line with $L_{S}=1$ and the point closest to $L_{A}=1/2$, respectively.  
Interestingly, both LE methods indicate the critical point $h_{N}=0.104$.

\begin{figure}[t]
	\centering
	\includegraphics[width=8.3cm]{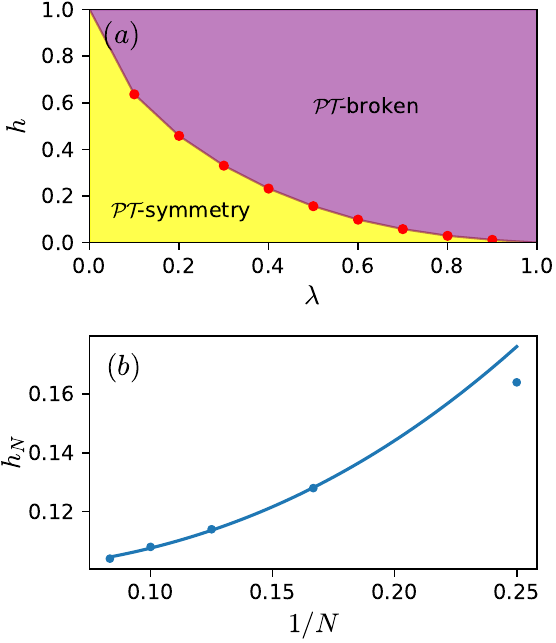}
	\caption{(a) Phase diagram of the Yang-Lee model as functions of $\lambda$ and $h$. 
	In this phase diagram, the red dot denotes the critical values $h_{\text{LE}}$ obtained from the LEs in the thermodynamic limit. The yellow region signifies the $\mathcal{PT}$-symmetric phase, while the purple region represents the $\mathcal{PT}$-symmetry broken phase. The solid line (served a guide for the eye) between these regions marks the Yang-Lee singularity edge. (b) Finite-size scaling of critical values $h_N$ with respect to the reciprocal of lattice size $N$ at $\lambda=0.6$. Here, the circles represent the critical values $h_{\text{N}}$ for finite lattice sizes, while the solid line is the fitting curve obtained using Eq.(\ref{eq:Hc}) with $\beta=12/5$, yielding the critical value shown in (a).} 
	\label{fig:phasescaling}
\end{figure}

However, we note that the critical point $h_{N}=0.104$ [see Appendix \ref{App:A}] for the $N=12$ lattice sites is slightly larger than the critical value $h_c = 0.0981$ in the thermodynamic limit \cite{von1991critical,wei2023tensor}.
The difference arises from the finite-size effect. To more accurately locate the Yang-Lee edge singularity, we perform simulations for both LEs with $N = 4, 6, 8, 10, 12$ by varying $h$ and $\lambda$.
The critical values for each $N$ are presented in Table \ref{tab:finiteLE} in Appendix \ref{App:A}.
The critical values, $h_{\text{LE}}$, in the thermodynamic limit are obtained from finite-size scaling, given by:
\begin{equation}
	h_N=h_{\text{LE}}+aN^{-\beta}
	\label{eq:Hc}
\end{equation}
where $\beta$ is the critical exponent, which is $\beta=12/5$ according to non-unitary CFT \cite{fisher1978yang,kortman1971density,cardy1985conformal,uzelac1979yang,von1991critical,shen2023proposal}.
The phase diagram of the Yang-Lee model is presented in Figs.\ref{fig:phasescaling}(a). Specifically, the numerical data and the fitting curve for $\lambda=0.6$ are presented in Figs.\ref{fig:phasescaling}(b), which are consistent with the analytical result from non-unitary CFT and other numerical results \cite{fisher1978yang,kortman1971density,cardy1985conformal,uzelac1979yang,von1991critical,bianchini2014entanglement,sanno2022engineering,yamada2022matrix,wei2023tensor,shen2023proposal}.

It is noteworthy that the critical values $h_{\text{LE}}$ derived from the LEs for $\lambda < 0.7$ more closely match the numerical critical points $h_c$ obtained from other methods, compared to the case where $\lambda > 0.7$ (see Table \ref{tab:infiniteLE}). This discrepancy arises because, for $\lambda > 0.7$, the critical point is nearer to the Ising transition, resulting in weaker non-Hermiticity. As a result, non-unitary time evolution is less sensitive in this regime compared to $\lambda < 0.7$. Additionally, it is observed that the associated-biorthogonal LE is more sensitive than the self-normal LE near the exceptional points.
We note that it is feasible to compute the self-normal LEs using the time-evolving block decimation method \cite{vidal2004efficient} beyond unitary evolution \cite{orus2008infinite}. However, numerically calculating the associated biorthogonal-biorthogonal LEs for large systems is more challenging, as determining the associated state requires the entire set of eigenvectors for constructing the projector.

Finally, we briefly discuss the experimental implementation of the self-normal LEs in the Yang-Lee model, following a protocol similar to the realization of dynamical quantum phase transitions in Hermitian systems \cite{jurcevic2017direct}. First, the initial state is prepared as the ground state of the initial Hamiltonian $H_0$ at $h = h_0$. Second, the system is suddenly switched to the Hamiltonian $H_f$ with a small change $\delta h$, and it evolves according to  Eq.(\ref{eq:timeEvolveR}). Third, the self-normal LEs, which represent the return probabilities to the ground state, are measured at discrete times. By varying the initial state with different values of $h_0$ and repeating these steps, experimental data can be obtained, as shown in Figs.\ref{fig:YLDynamics}.
The properties of the Yang-Lee edge singularity are then analyzed using finite-size scaling theory, as demonstrated in Figs.\ref{fig:phasescaling}(b). 

\section {Conclusion}
\label{sec:conclusion}
In summary, we have explored the concepts of both the self-normal LE and the associated-biorthogonal LE to analyze phase transitions in non-Hermitian systems. Our investigation reveals that LEs can exhibit fundamentally different behaviors in both the real-energy and complex-energy regimes. This distinction, particularly near the exceptional point, facilitates the detection of phase transitions and critical exponents. Our findings highlight that LEs serve as excellent indicators for describing the Yang-Lee edge singularity, aligning well with other theoretical predictions.

Furthermore, we have demonstrated that the NHTI model lacks finite-size scaling, whereas the Yang-Lee models exhibit pronounced finite-size scaling. This difference suggests that the exceptional points in the NHTI model and the Yang-Lee models are distinct. 
Although we show that Yang-Lee edge singularities can be studied as exceptional points, the one-to-one correspondence between many-body exceptional points and Yang-Lee edge singularities is still not well established. Exploring a general correspondence between exceptional points and Yang-Lee edge singularities would be a highly interesting avenue for future research.
In addition, the universality class of the exceptional points in non-Hermitian many-body systems would be another interesting topic for future study.

{\it Note added.-} During the review process of our work, we became aware of a related study \cite{xu2024characterizing} that explores the correspondence between exceptional points and Yang-Lee edge singularities, based on the quantum-classical correspondence.

\begin{acknowledgments}
G.S. is appreciative of support from "the Fundamental Research Funds for the Central Universities under the Grant No. NS2023055", the NSFC under the Grant No. 11704186, and the High Performance Computing Platform of Nanjing University of Aeronautics and Astronautics. 
M.L. and S.S. are appreciative of support from the Innovation Training Fund for College Students under Grant No. 2023CX021023.
\end{acknowledgments}

\bibliography{YLref}

\clearpage
\begin{widetext}
\appendix
\section{Numerical data for finite systems}
\renewcommand{\thetable}{\Alph{section}\arabic{table}}
\setcounter{table}{0}
\label{App:A}
In this appendix, we present the detailed numerical results of the critical points $h_{\text{N}}$ of the Yang-Lee model for finite systems, as calculated from the LEs. 
The data are obtained by varying the system size $N$ from 6 to 12.
\begin{table}[h]
    \renewcommand{\arraystretch}{1.1}
    \setlength{\tabcolsep}{8pt}
    \centering
    \caption{Critical points $h_{N}$ with respect to $\lambda$, ranging from $0.1$ to $0.9$. }
    \label{tab:finiteLE}
    \begin{tabular}{|c|c|c|c|c|c|}
    \hline
        $\lambda$ & $h_{\text{N=6}}$ & $h_{\text{N=8}}$ & $h_{\text{N=10}}$ & $h_{\text{N=12}}$ \\ \hline
        0.1 & 0.642 & 0.639 & 0.638 & 0.637 \\ \hline
        0.2 & 0.469 & 0.464 & 0.461 & 0.460 \\ \hline
        0.3 & 0.345 & 0.338 & 0.334 & 0.333 \\ \hline
        0.4 & 0.252 & 0.242 & 0.237 & 0.236 \\ \hline
        0.5 & 0.181 & 0.169 & 0.163 & 0.161 \\ \hline
        0.6 & 0.128 & 0.114 & 0.107 & 0.104 \\ \hline
        0.7 & 0.088 & 0.074 & 0.067 & 0.063 \\ \hline
        0.8 & 0.060 & 0.046 & 0.039 & 0.034 \\ \hline
        0.9 & 0.041 & 0.027 & 0.021 & 0.017 \\ 
        \hline
    \end{tabular}
\end{table}

\end{widetext}

\end{document}